\documentclass[aps,prd,superscriptaddress,amsfonts,amssymb,amsmath,eqsecnum,nofootinbib,twocolumn,floatfix]{revtex4}
  \usepackage{color,graphicx}
  \usepackage[utf8]{inputenc}
  \usepackage[T2A]{fontenc}
  \usepackage[english]{babel}
  \definecolor{darkblue}{rgb}{0,0,0.7}
 \definecolor{darkred}{rgb}{0.7,0,0}
 \definecolor{darkgreen}{rgb}{0,0.4,0}
 \usepackage[unicode, colorlinks, citecolor=darkblue, linkcolor=darkred, urlcolor=blue]{hyperref}
  \newcommand{\vy}[1]{{\color{darkblue}{ #1}}}

  \newcommand{\na}[1]{{\color{darkgreen}   #1 }}
 \allowdisplaybreaks
\begin{document}

\author{Sergey P. Vyatchanin}

\affiliation{Faculty of Physics, M.V. Lomonosov Moscow State University, Leninskie Gory, Moscow 119991, Russia}
\affiliation{Quantum Technology Centre, M.V. Lomonosov Moscow State University, Leninskie Gory, Moscow 119991, Russia}

\author{Albert I. Nazmiev}
\affiliation{Faculty of Physics, M.V. Lomonosov Moscow State University, Leninskie Gory, Moscow 119991, Russia}

\author{Andrey B. Matsko}

\affiliation{Jet Propulsion Laboratory, California Institute of Technology, 4800 Oak Grove Drive, Pasadena, California 91109-8099, USA}

\date{\today}
	
\title{Broadband Dichromatic Variational Measurement}

\begin{abstract}
Standard Quantum Limit (SQL) of a classical mechanical force detection results from quantum back action impinged by the meter on a probe mechanical transducer perturbed by the force of interest. In this paper we introduce a technique of continuous \vy{broadband} back action avoiding measurements for the case of a resonant signal force acting on a linear mechanical oscillator supporting one of mirrors of an optical Michelson-Sagnac Interferometer (MSI). The interferometer with the movable mirror is an opto-mechanical transducer able to support polychromatic probe field. The method involves a dichromatic optical probe resonant with the MSI modes and having frequency separation equal to the mechanical frequency. We show that analyzing each of the harmonics of the probe reflected from the mechanical system separately and postprocessing the measurement results allows excluding the back action in a broad frequency band and measuring the force with sensitivity better than SQL.
\end{abstract}

\maketitle

\section{Introduction}

Optical transducers are frequently used for observation of mechanical motion. They allow measuring displacement, speed, acceleration, and rotation of mechanical systems. Mechanical motion can change frequency, amplitude and phase of the probe light, that is processed to obtain information about the motion. The measurement sensitivity can be extremely high. For instance, a relative mechanical displacement orders of magnitude smaller than a proton size can be detected. This feature is utilized in gravitational wave detectors \cite{aLIGO2013,aLIGO2015,MartynovPRD16,AserneseCQG15, DooleyCQG16,AsoPRD13}, in magnetometers \cite{ForstnerPRL2012, LiOptica2018}, and in torque sensors \cite{WuPRX2014, KimNC2016, AhnNT2020}. 

The fundamental sensitivity limitations of the measurement always were of interest. One of the limits results from the fundamental thermodynamic fluctuations of the probe mechanical system. The absolute position detection is restricted due to the Nyquist noise. However, this obstacle can be removed if one intends to measure a variation of the position, not its absolute value. The thermal noise does not limit sensitivity of the measurement that occurs much faster than the system ringdown time.

Another limitation comes from the quantum noise of the meter. On one hand, the accuracy of the measurements of the observables of the meter is restricted because of their fundamental quantum fluctuations, represented by the shot noise for the optical probe wave. On the other hand, the sensitivity is impacted by the perturbation of the state of the probe mass due to the mechanical action of the meter on the mass. This effect is called ``quantum back action''. In the case of optical meter the mechanical perturbation results from the pressure of light. Interplay between these two phenomena results in so called standard quantum limit (SQL) \cite{Braginsky68,BrKh92} of the measurement sensitivity. 

The value of the SQL depends on the measurement system as well as the measurement strategy and measurement observable. In the case of the detection of a classical force acting on a mechanical probe mass, the SQL can be avoided in a configuration supporting opto-mechanical velocity measurement  \cite{90BrKhPLA,00a1BrGoKhThPRD}. The limit also can be surpassed using opto-mechanical rigidity \cite{99a1BrKhPLA, 01a1KhPLA}. Preparation of the probe light in a nonclassical state \cite{LigoNatPh11, LigoNatPhot13, TsePRL19, AsernesePRL19, YapNatPhot20, YuNature20, CripeNat19} as well as detection of a variation of a strongly perturbed optical quadrature \cite{93a1VyMaJETP, 95a1VyZuPLA, 02a1KiLeMaThVyPRD} curbs the quantum back action and lifts SQL. The back action and SQL can avoided with coherent quantum noise cancellation \cite{TsangPRL2010, PolzikAdPh2014, MollerNature2017} as well as compensation using an auxiliary medium with negative nonlinearity \cite{matsko99prl}.

Optimization of the detection scheme by utilizing a few optical frequency harmonics as a probe allows beating the SQL of a force detection. The technique was introduced forty years ago  \cite{80a1BrThVo, 81a1BrVoKh} and then expanded to various measurement configurations \cite{Clerk08, Clerk2015, Pirkkalainen2015, Buchmann2016, 18a1VyMaJOSA}. Additionally, usage of a dichromatic optical probe in a resonant optical transducer may lead to observation of such phenomena as negative radiation pressure \cite{Povinelli05ol,maslov13pra}, optical quadrature-dependent quantum back action \cite{93a1VyMaJETP,96a1VyMaJETP,matsko97apb,02a1KiLeMaThVyPRD} as well as mechanical velocity-dependent interaction \cite{90BrKhPLA,00a1BrGoKhThPRD}. All these phenomena are useful for back action suppression. 
 
Noncommutativity between the probe noise and the quantum back action noise is the reason for SQL. In a simple displacement sensor the probe noise is represented by the phase noise of light and the back action noise depends on the amplitude noise of light. The signal is contained in the phase of the probe. The relative phase noise decreases with optical power. The relative back action noise increases with the power. The signal to noise ratio optimizes at a specific power value. The optimal measurement sensitivity corresponds to SQL. Because phase and amplitude quantum fluctuations of the same wave do not commute, it is not possible to measure the amplitude noise and subtract it from the measurement result. 

In this paper we suggest a measurement procedure that allows post-processing of the measurement result and subtracting the quantum back action contribution. The measurement takes advantage of usage of a dichromatic optical probe in a resonant displacement transducer. We show that the signal results in modification of the power distribution between the spectral components of the probe, while the back action is proportional to the total power of the probe harmonics. Fundamentally, the power difference and total power can be measured independently. As the result, the quantum back action can be removed from the measurement result by post-processing of the measurement results.

In the measurement procedure the probe mass is strongly perturbed. In this way the technique has similar spirit with the variational approach leading to accurate measurement of a variation of a quantum systems in spite of the strong perturbation of the system parameters. Unlike standard variational measurement, the described here technique is broadband. The back action can be removed at all spectral frequencies. 

The proposed here technique is especially efficient when the signal force is resonant with the mechanical probe mass suspension. In this regime the external force modifies the power redistribution between the probe spectral components most efficiently. The variational measurement technique based on a monochromatic probe light does not perform well in this case.

The paper is organized as follows. The physical model of the measurement system is introduced in Section II.  The mean amplitudes of the system parameters, the quantum fluctuations and associated noise components are studied in Section III. The optimal sensitivity of the measurement is also found in this Section. The impact of the parasitic sidebands is analyzed in Section IV.  Section V concludes the paper.

\section{Physical Model}

\begin{figure}
\includegraphics[width=0.35\textwidth]{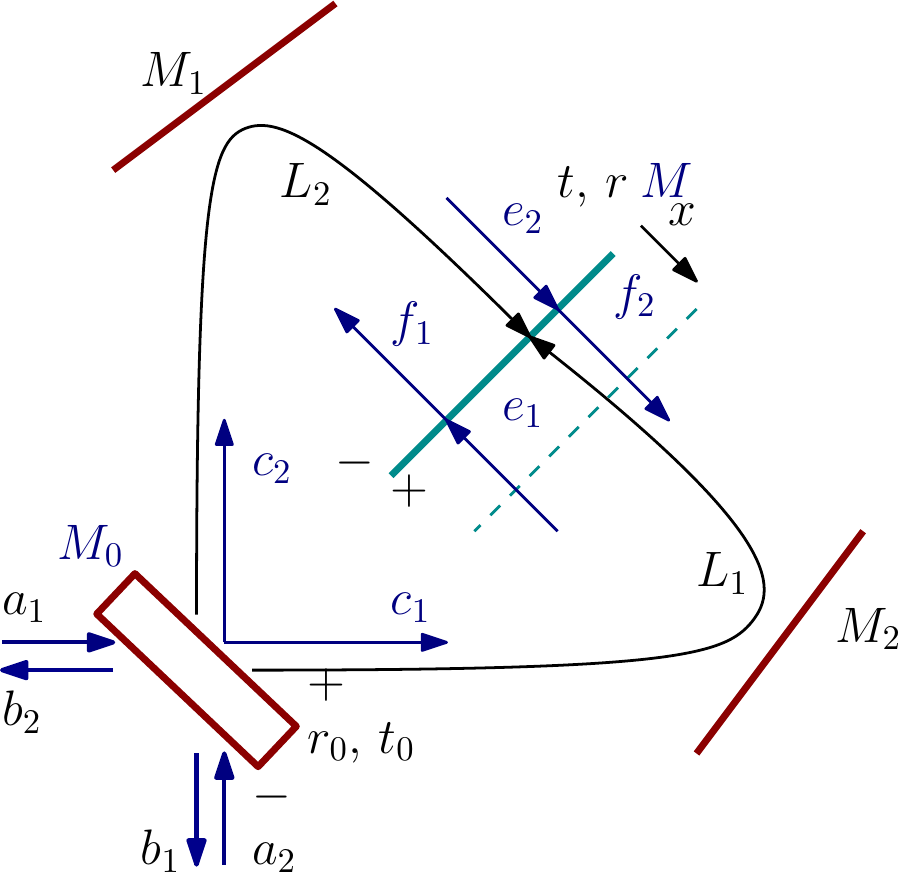}
 \caption{High-finesse ring cavity with  input mirror $M_0$ with transmission $t_0$ and reflectivity $r_0$ ($t_0\ll r_0$) and dielectric mirror $M$ (membrane) inside with transmission $t$ and reflectivity $r$ ($r\ll t$).}\label{TriCav}
\end{figure}

Let us consider an opto-mechanical system consisting of two externally pumped optical modes coupled with each other and with a mechanical oscillator (Fig.~\ref{TriCav}). The difference between the optical mode frequencies is equal to the frequency of the mechanical oscillator. In what follows we show feasibility of the broadband detection of a small signal force acting on the mechanical oscillator while keeping sensitivity of the measurement better than the SQL.

The opto-mechanical system can be realized in a ring cavity with coupled clockwise and counterclockwise modes. The uncoupled modes are frequency degenerate. Let us assume that their frequency is equal to $\omega_0$. Introducing a membrane inside the cavity (see Fig.~\ref{TriCav}) removes degeneracy and frequencies of the modes split (so called coherent coupling \cite{LiPRA2019})
\begin{align}
\label{omegaLH}
   \omega_- & =\omega_0 - |\kappa|, \quad \omega_+  =\omega_0 + |\kappa|, 
\end{align}
where splitting factor $\kappa$  depends on transparency  of the membrane as well as on its position. Parameter $\kappa$ is a complex value. Specific feature of coherent coupling is such that the absolute value of $|\kappa|$ depends on optical parameters of the membrane and phase of $\kappa$ depends on the membrane position \cite{LiPRA2019}. 

\begin{figure}
 \includegraphics[width=0.45\textwidth]{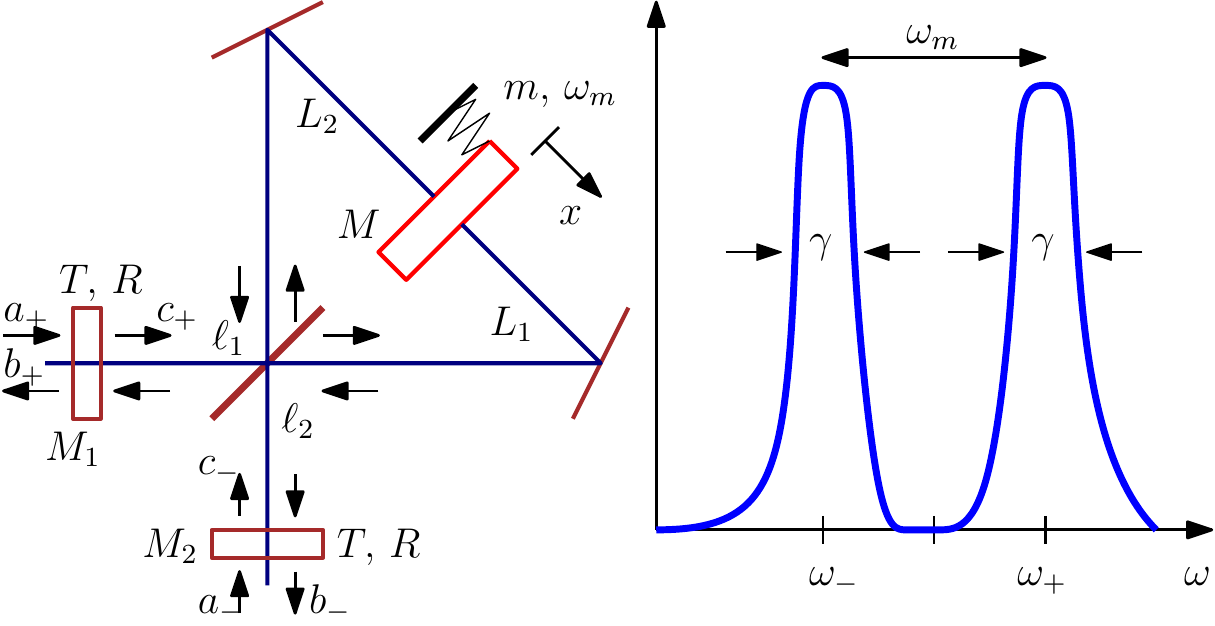}
 \caption{Schematic of the Michelson-Sagnac interferometer in which mirror $M$ is completely reflecting. The mirror is a test mass $m$ of the mechanical oscillator with frequency $\omega_m$. Two eigenmodes with frequencies $\omega_-,\ \omega_+$ are coupled with the mechanical oscillator. Relaxation rate $\gamma$ is the same for the both modes, $\gamma\ll \omega_m$. Modes $\omega_\pm$ are resonantly pumped.}\label{Cav2modes}
\end{figure}

Another example of a scheme enabling the back action evading measurement is the Michelson-Sagnac interferometer (MSI) shown in Fig.~\ref{Cav2modes}. The system also has two degenerate modes. If the position of a perfectly reflecting mirror $M$ is fixed, one MSI mode, characterized with frequency $\omega_+$, is given by a light wave which travels between $M_1$ and BS. The light is split on the BS and after reflection from mirror $M$ returns exactly to $M_1$. It does not propagate to $M_2$. The other mode, characterized with frequency $\omega_-$,  is represented by a wave which travels from $M_2$ to BS and after reflection from $M$ returns to $M_2$ and does not propagate to $M_1$. The frequencies of the modes, $\omega_\pm$, are controlled by variation of path distances $\ell_1,\ \ell_2$. Variation of the position of mirror $M$ provides coupling between the modes.  Mirror $M$ is a test mass of the mechanical oscillator with mass $m$ and eigenfrequency $\omega_m$. 

\subsection{Main assumptions}

Let us consider scheme shown in Fig. 2. We  assume that the relaxation rates of the eigen modes are identical and characterized with the full width at the half maximum (FWHM) equal to $2\gamma$.  The  mechanical relaxation rate is small if compared with the optical one: $ \gamma_m\ll \gamma$. We also assume that the conditions of the resolved side band interaction and frequency synchronisation are valid:  
\begin{align}
 \label{RSB}
  \gamma_m\ll \gamma \ll \omega_m, \quad \omega_+-\omega_- =\omega_m
\end{align}
The resonance curves shown in the right panel of Fig.~\ref{Cav2modes} illustrate the conditions accepted above.

For the sake of simplicity we also assume that following conditions are valid
\begin{align}
 \label{condLg}
 L_1 &\simeq L_2,\quad L_1,\ L_2 \gg \ell_1,\ \ell_2.
\end{align}

\subsection{Hamiltonian}

The generalized Hamiltonian describing the system can be presented in form
\begin{subequations}
   \label{Halt}
  \begin{align}
  H  & = H_0 + H_\text{int}+H_T+H_\gamma+H_{T, \, m}+H_{\gamma_ m} ,\\
  \label{H0}
  H_0 &=\hslash \omega_+c_+^\dag c_+  + \hslash \omega_-c_-^\dag c_-
         +\hslash \omega_m d^\dag d,\\
  \label{Hint}
      H_\text{int} & = \frac{\hslash }{i}
      \left(\eta c_+^\dag c_-d - \eta^* c_+ c_-^\dag d^\dag \right).
  \end{align}
  \end{subequations}
$H_\text{int}$ is the Hamiltonian of the interaction between modes,  $d,\ d^\dag$ are annihilation and creation operators of the mechanical oscillator,  $c_\pm,\ c^\dag_\pm$ are annihilation and creation operators of the corresponding optical modes. The operator of coordinate $x$ of the mechanical oscillator can be presented in form
\begin{align}
\label{x}
 x= x_0\left(d + d^\dag\right),\quad x_0=\sqrt\frac{\hslash}{2m \omega_m}.
\end{align}
The coupling constant $\eta$ can be written as
\begin{align}
 \label{eta}
 |\eta| &\simeq \frac{x_0}{L}\, \omega_0,\quad L\simeq L_1,\, L_2,\quad \omega_0\simeq \omega_\pm
\end{align}
$H_T$ is the Hamiltonian describing the environment (thermal bath) and  $H_\gamma$ is the Hamiltonian of the coupling between the environment and the optical modes resulting in decay rate $\gamma$.
Similarly, $H_{T, \, m}$ is the Hamiltonian of the environment and $H_{\gamma_m}$ is the Hamiltonian describing coupling between the environment and the mechanical oscillator resulting in decay rate $\gamma_m$. See Appendix~\ref{IntrDeriv} for details.

It is convenient to separate the expectation values of the wave amplitudes as well as their fluctuation parts and assume that the fluctuations are small:
 \begin{align}
  \label{expA}
  \mathcal A_\pm & = \left(A_\pm + \hat a_\pm\right)e^{-i\omega_\pm t},\\
  \mathcal B_\pm  & =\left( B_\pm + \hat b_\pm \right)e^{-i\omega_\pm t}.
 \end{align}
$A_\pm$ and $B_\pm$ stand for the expectation values of the amplitudes of the corresponding optical waves and $a_\pm$ and $b_\pm$ represent the fluctuations, $|A_\pm|^2 \gg \langle a_\pm^\dag a_\pm \rangle$ and $|B_\pm|^2 \gg \langle b_\pm^\dag b_\pm \rangle$, where $\langle \dots \rangle$ stands for ensemble averaging.

The normalization of the amplitudes is selected so that $\hslash \omega_\pm |A_\pm|^2$ describes the optical power \cite{02a1KiLeMaThVyPRD}. We also consider only spectral components around carrier frequencies $\omega_\pm$ and drop the harmonics centered at frequencies $(\omega_++\omega_m)$ and $(\omega_--\omega_m)$ far from the corresponding resonances.

The Hamiltonian of the system allows us to write the equations of motion for the intracavity fields. The complete derivation of these equations is presented in Appendix \ref{IntrDeriv}. 
\begin{subequations}
\begin{align}
 \label{c+T}
  \hat {\dot   c}_+  + \gamma \hat c_+ + \eta c_- \hat d &= \sqrt {2 \gamma} \hat a_+,\\
  \label{c-T}
  \hat {\dot   c}_-  + \gamma \hat c_- - \eta c_+ \hat d^\dag &= \sqrt {2 \gamma} \hat a_-.
\end{align}
The input-output relations are
\begin{align}
 \label{outputT}
 \hat b_\pm= -\hat a_\pm + \sqrt{2\gamma} \hat c_\pm
\end{align} 
\end{subequations}

\section{Solution}

Using the Hamiltonian formalism we obtain set of equations for the expectation values
\begin{subequations}
 \begin{align}
 \label{C+}
  \gamma C_+ +\eta C_-D &= \sqrt{2\gamma}\, A_+,\\
  \label{C-}
  \gamma C_- -\eta^* C_+D^* &= \sqrt{2\gamma}\, A_-,\\
  \label{D}
  \gamma_m D &=\eta^* C_+ C_-^*,\\
  \label{D*}
  \gamma_m D^* &=\eta C_+^* C_-
 \end{align}
 \end{subequations}

Introducing parameter $\nu=|\eta|^2/\gamma\gamma_m$ we arrive at
\begin{subequations}
\begin{align}
  C_+ \left( 1 + \frac{\nu g^2 |A_-|^2}{\left( 1 -\nu |C_+|^2\right)^2} \right) & = \sqrt \frac{2}{\gamma}\, A_+,\\
  C_- \left( 1 - \frac{\nu g^2 |A_+|^2}{\left( 1 +\nu |C_-|^2\right)^2} \right) & = \sqrt \frac{2}{\gamma}\, A_-
 \end{align}
  \end{subequations}
The amplitudes $C_+,\ C_-$ are limited due to the ponderomotive nonlinearity. The classical resonant ponderomotive force creates mechanical oscillations with amplitude $D$ which can be large. These oscillations are classical and can be suppressed using a regular force optimized for the known amplitudes of the probe fields. We also can use two orthogonal polarizations in the scheme shown in Fig.~\ref{Cav2modes} to reduce the undesirable resonant excitation of the mechanical oscillator (see Appendix \ref{AppD} for details).  In what follows we omit them from consideration and assume that $D=0$ and $C_\pm = \sqrt{2/\gamma}\, A_\pm$.

\subsection{Langevin equations}

The equations of motion for the Fourier amplitudes of the operators describing the intracavity field as well as mechanical oscillator, $c_\pm$ and $d$, can be written in form using (\ref{c+T}, \ref{c-T}):
\begin{subequations}
 \label{set1}
\begin{align} 
\label{c+}   
 (\gamma - i\Omega)c_+(\Omega) & +  \eta C_- d(\Omega) = \sqrt {2 \gamma}  { a}_{+}(\Omega),\\
\label{c-} 
 (\gamma - i\Omega) c_-(\Omega) & - \eta^* C_+ d^\dag(-\Omega)  = \sqrt {2\gamma} a_-(\Omega),\\
\label{dd}
 (\gamma_m - i\Omega) d(\Omega) -  &    \eta^*  \Big(C_-^*c_+(\Omega)+C_+c_-^\dag(-\Omega)\Big) =\\
 \nonumber
    & =  \sqrt{2\gamma_m}\, q(\Omega)+     i f_s(\Omega)\\ 
\label{output}
  b_{ \pm}(\Omega) &= - { a}_{\pm}(\Omega) + \sqrt {2 \gamma}\, { c}_\pm(\Omega),
\end{align} 
\end{subequations}
where  $b_\pm$ are the output Fourier amplitudes of the optical waves. The incident waves are in the coherent state, so the operators $\hat a_\pm$ are characterized with the following commutators and correlators
\begin{align}
  \label{comm}
  \left[\hat a_\pm(t), \hat a_\pm^\dag(t')\right] &=   \delta(t-t'),\\
  \label{corr}
  \left\langle\hat a_\pm(t) \hat a_\pm^\dag(t')\right\rangle &= \delta(t-t'),
\end{align}
$\langle \dots \rangle$ stands for ensemble averaging.

The Fourier transform of these operators are introduced as follows
\begin{align}
 \hat a_\pm (t) &= \int_{-\infty}^\infty a_\pm(\Omega) \, e^{-i\Omega t}\, \frac{d\Omega}{2\pi}.
\end{align}
Same is true for the other operators.
Using \eqref{comm} and \eqref{corr} we derive commutators and correlators for the Fourier amplitudes of the input fluctuation operators
 \begin{align}
  \label{comm1}
  \left[ a_\pm(\Omega),  a_\pm^\dag(\Omega')\right] &= 2\pi\,\delta(\Omega -\Omega'),\\
  \label{corr1}
  \left\langle a_\pm(\Omega)  a_\pm^\dag(\Omega')\right\rangle &= 2\pi\, \delta(\Omega -\Omega')
\end{align}

For the signal force we also introduce its Fourier amplitude assuming that the force is the resonant one acting during time interval $\tau$:
\begin{align}
\label{Fs}
F_S(t)&= F_{s0}\cos(\omega_m t + \psi_f) = \\
    = &F_{s}(t) e^{-i\omega_m t} + F_{s}^*(t) e^{i\omega_m t},\quad 
  -\frac{\tau}{2} < t < \frac{\tau}{2},\nonumber\\
  \label{fs}
  f_s(\Omega)& = \frac{F_s(\Omega)}{\sqrt{2\hslash \omega_m m}},\quad 
  f_{s0} = \frac{F_{s0}(\Omega)}{\sqrt{2\hslash \omega_m m}}= 2f_s.
\end{align}
where  $F_s(\Omega)$ is the Fourier amplitude of the slow complex amplitude $F_s(t)$. In general case $F_s(\Omega)\ne F_s^*(-\Omega)$. 

The thermal mechanical noise operators are described using expressions
\begin{subequations}
 \label{qDef}
\begin{align}
 \label{commq}
  \left[ q(\Omega),\,  q^\dag(\Omega')\right] &= 2\pi\,  \delta(\Omega-\Omega'),\\
  \label{corrq}
  \left\langle q(\Omega)\,  q^\dag(\Omega')\right\rangle 
    &= \big(2n_T+1\big)\,2\pi\,  \delta(\Omega-\Omega'),\\
    n_T&= \dfrac{\hslash \omega_m}{1 - e^{-\hslash \omega_m/\kappa_BT} }.
\end{align}
\end{subequations}
Here $\kappa_B$ is Boltzmann constant, $T$ is the ambient  temperature.

\subsection{Solution of the Langevin equations}

For the sake of simplicity we assume that the phases of the probe harmonics are selected so that
\begin{align}
 \label{real}
 C_+=C_+^*= C_-=C_-^* =C, \quad \eta =\eta^*
\end{align}
Introducing quadrature amplitudes 
\begin{subequations}
\label{quadDef}
 \begin{align}
  a_{\pm a} &= \frac{a_\pm (\Omega) +a_\pm ^\dag(-\Omega)}{\sqrt 2}\,,\\
	 a_{\pm \phi} &= \frac{a_\pm (\Omega) -a_\pm ^\dag(-\Omega)}{i\sqrt 2}\,.
 \end{align}
\end{subequations}
and using \eqref{set1} we obtain
\begin{subequations}
  \label{quadIn}
 \begin{align}
  \label{ca+}   
 (\gamma - i\Omega)c_{+a} +  \eta C  d_a &= \sqrt {2 \gamma}  { a}_{ +a},\\
 \label{cphi+}   
 (\gamma - i\Omega)c_{+\phi} +  \eta C  d_\phi &= \sqrt {2 \gamma}  { a}_{ +\phi},\\
\label{ca-} 
 (\gamma - i\Omega) c_{-a} - \eta C  d_a &= \sqrt {2 \gamma}  { a}_{-a},\\
\label{cphi-} 
 (\gamma - i\Omega) c_{-\phi} + \eta  C  d_\phi &= \sqrt {2 \gamma}  { a}_{-\phi},\\
\label{da}
 (\gamma_m - i\Omega)  d_a &- \eta C \Big(c_{+a}+c_{-a}\Big) =\\
        &= \sqrt {2 \gamma_m} q_a -  f_{s\,\phi},\\
\label{dphi}
 (\gamma_m - i\Omega)  d_\phi & - \eta C\Big(c_{+\phi} - c_{-\phi}\Big) =\\
        &= \sqrt {2 \gamma_m} q_{\phi} +  f_{s\,a}.
 \end{align}
 \end{subequations}
Please note that sum $(c_{+a} +c_{-a})$ does not contain  information on the mechanical motion (term $\sim d_a$ is absent), but produces the back action term in \eqref{da}.

Introducing
\begin{align}
 \label{gDef}
 g_{a\pm} &= \frac{c_{+a}\pm c_{-a}}{\sqrt 2},\quad 
  g_{\phi\pm} = \frac{c_{+\phi}\pm c_{-\phi}}{\sqrt 2},\\
 \label{alphaDef}
 \alpha_{a\pm}&= \frac{a_{+a}\pm a_{-a}}{\sqrt 2}\,,\quad
    \alpha_{\phi\pm}= \frac{a_{+\phi}\pm a_{-\phi}}{\sqrt 2},\\
 \label{betaDef}
 \beta_{a\pm}&= \frac{b_{+a}\pm b_{-a}}{\sqrt 2}\,,\quad
    \beta_{\phi\pm}= \frac{b_{+\phi}\pm b_{-\phi}}{\sqrt 2}
\end{align}
and rewriting \eqref{quadIn} in the new notations
\begin{subequations}
  \label{quadIn2}
 \begin{align}
  \label{ga+}   
 (\gamma - i\Omega)g_{a+} = \sqrt {2 \gamma}  \alpha_{ a+},\\
 \label{ga-} 
 (\gamma - i\Omega) g_{a-} + \sqrt 2 \eta C  d_a &= \sqrt {2 \gamma}  \alpha_{a-},\\
 \label{da2}
 (\gamma_m - i\Omega)  d_a - \sqrt 2 \eta C g_{a+} 
        &= \sqrt {2 \gamma_m} q_a -  f_{s\,\phi},\\
 \label{gphi+}   
 (\gamma - i\Omega)g_{\phi+} +  \sqrt 2\eta C  d_\phi &= \sqrt {2 \gamma}  \alpha_{ \phi+},\\
\label{gphi-} 
 (\gamma - i\Omega) g_{-\phi}  &= \sqrt {2 \gamma} \alpha_{\phi-},\\
\label{dphi2}
(\gamma_m - i\Omega)  d_\phi  - \sqrt 2 \eta Cg_{\phi-} 
        &= \sqrt {2 \gamma_m} q_{\phi} +  f_{s\,a}.
 \end{align}
 \end{subequations}
we find that sets (\ref{ga+}, \ref{ga-}, \ref{da2}) and (\ref{gphi+}, \ref{gphi-}, \ref{dphi2}) are decoupled.

It is convenient to present the solution of set (\ref{ga+}, \ref{ga-}, \ref{da2}) for the amplitude quadratures in form
\begin{subequations}
 \label{betaMSIa}
 \begin{align}
  \beta_{+a}   &= \xi\, \alpha_{+a},\quad \xi=\frac{\gamma+i\Omega}{\gamma-i\Omega},\\
  \label{beta-a}
  \beta_{-a}  &=\xi\left(\alpha_{-a} - \frac{\mathcal K\, \alpha_{+a}}{\gamma_m-i\Omega}\right)-\\
    &\qquad         - \frac{\sqrt{\xi \mathcal K }}{\gamma_m-i\Omega}
            \left(\sqrt {2 \gamma_m} q_a - f_{s\,\phi}\right),\\
        &\quad  \mathcal K\equiv \frac{4  \gamma\,\eta^2 C^2}{\gamma^2+\Omega^2}
 \end{align}
 \end{subequations}
As expected, in Eq.~\eqref{beta-a} the back action term is proportional to the normalized probe power $\mathcal K$. However, this term can be excluded in the post processing.
One  can measure {\em both}  $\beta_{+a}$ and $\beta_{-a}$ simultaneously and subtract $\beta_{+a}$ from  $\beta_{-a}$ to remove the back action completely. It means that we can measure combination
 \begin{align}
  \beta_{-a}^\text{comb} &= \beta_{-a} + \xi \, \frac{\mathcal K\, \alpha_{+a}}{\gamma_m-i\Omega} =\\
  \label{beta-a2}
         &=\xi\alpha_{-a} - \frac{\sqrt{\xi \mathcal K }}{\gamma_m-i\Omega}
            \left(\sqrt {2 \gamma_m} q_a - f_{s\,\phi}\right),
 \end{align}
 which is completely back action free (!). This is the main finding of the study.
 
Let us write the force detection condition in terms of single-sided power spectral density $S_{f}(\Omega)$ recalculated to signal force \eqref{Fs}. Demanding signal-to-noise ratio to exceed unity we get
 \begin{align}
  \label{SfDef}
  f_{s0} \ge \sqrt{S_{f}(\Omega)\cdot \frac{\Delta\Omega}{2\pi} },
 \end{align}
 where $\Delta \Omega \simeq 2\pi/\tau$. 
 Using (\ref{corr1}, \ref{corrq}) we obtain for the case when we measure  $ \beta_{-a}$ \eqref{beta-a}
 \begin{align}
  \label{Sf}
  S_{f}(\Omega) &= 2\gamma_m\big(2n_T+1\big) 
    + \frac{\big(\gamma_m^2+\Omega^2\big)}{|\mathcal K|} +|\mathcal K|\ge\\
    \label{SQL}
    &\ge 2\gamma_m\big(2n_T+1\big) +S_{SQL,f},\\ 
  S_{SQL,f} &= 2\sqrt{\gamma_m^2+\Omega^2}
 \end{align}
The sensitivity is restricted by SQL.
If me measure $\beta_{-a}^\text{comb}$ \eqref{beta-a2} the spectral density is not limited by SQL
 \begin{align}
  \label{Sfb}
  S_{f}(\Omega) &= 2\gamma_m\big(2n_T+1\big) + \frac{\big(\gamma_m^2+\Omega^2\big)}{|\mathcal K|}
 \end{align}
Here the first term describes thermal noise and the last one stands for the quantum measurement noise (shot noise). The quantum measurement noise decreases with the power increase. The back action term is absent. 

It worth noting that the thermal noise term is present in any opto-mechanical detection scheme. It can exceed the measurement error related to the measurement apparatus rather significantly. However, a proper measurement procedure allows to suppress this noise and also exclude the initial quantum uncertainty associated with the mechanical system. The main requirement for such a measurement is fast interrogation time, which should be much shorter than the ring down time of the mechanical system. This is possible if the measurement bandwidth exceeds the bandwidth of the mechanical mode. Sensitivity of narrowband resonant measurements is usually limited by the thermal noise.
 
Instead of the amplitude quadratures one can measure sum and differences of the phase quadratures. Solving set (\ref{gphi+}, \ref{gphi-}, \ref{dphi2}) we arrive at
 \begin{subequations}
  \label{betaMSIphi}
 \begin{align}
  \beta_{-\phi} &= \xi\, \alpha_{-\phi},\\
  \label{beta-phi}
  \beta_{+\phi}  &=\xi\left(\alpha_{+\phi} - \frac{\mathcal K\, \alpha_{-\phi}}{\gamma_m-i\Omega}\right)-\\
    &\qquad         - \frac{\sqrt{\xi \mathcal K }}{\gamma_m-i\Omega}
            \left(\sqrt {2 \gamma_m} q_\phi +  f_{s\,a}\right).
 \end{align}
\end{subequations}
We can measure quadratures $\beta_{\pm\phi}$ simultaneously and subtract back action proportional to $\beta_{-\phi}$ from $\beta_{+\phi}$.

Finally, a generalization is possible for a pair of quadratures with arbitrary parameter $\varphi$
\begin{subequations}
 \label{quadphi}
\begin{align}
 b_{+\varphi} &= b_{+a}\cos\varphi +b_{+\phi}\sin\varphi,\\
   b_{-\varphi}&= b_{-a}\cos\varphi -b_{-\phi}\sin\varphi
\end{align}
\end{subequations}
The sum $b_{+\varphi}+ b_{-\varphi}$ is not disturbed by the mechanical motion but contains the term proportional to the back action force, whereas the difference $b_{+\varphi}- b_{-\varphi}$ contains the term proportional to mechanical motion (with back action and signal). The back action term can be measured and subtracted from the force measurement result.

\begin{figure}
 \includegraphics[width=0.45\textwidth]{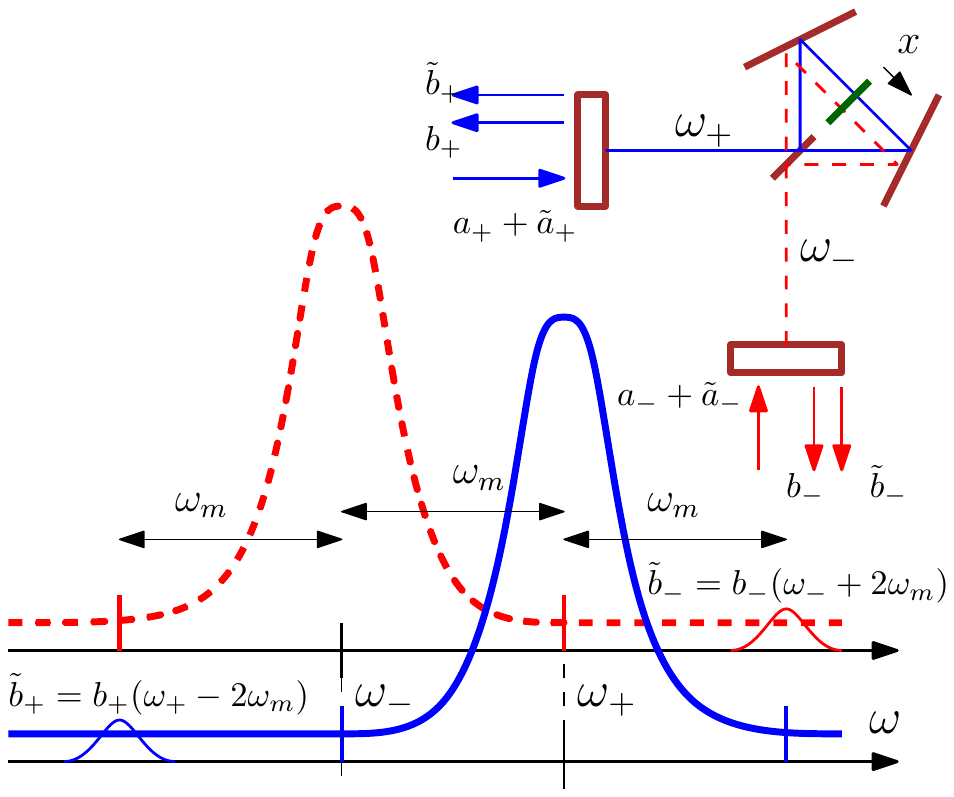}
 \caption{Side bands $\tilde c_\pm =c_\pm(\omega_\pm \mp 2\omega_m)$ inside cavity produce parasitic back action force acting on mechanical oscillator. We assume that one can measure waves $b_\pm$ and side bands $\tilde b_\pm$ separately. }\label{wings2}
 \end{figure}

\section{Parasitic back action}

 Fluctuation force acting on the mechanical oscillator is proportional to cross product $(c_-c_+^\dag + c_-^\dag c_+)$ of the probe modes. In means that the fluctuation fields characterized with Fourier amplitude $\tilde c_-(\Omega)= c_-(\omega_-+2\omega_m +\Omega)$ centered at frequencies in the vicinity of $\omega_-+2\omega_m$ and Fourier amplitude $\tilde c_+(\Omega) = c_+(\omega_+-2\omega_m +\Omega)$ centered at frequency in the vicinity of $\omega_+-2\omega_m$ (see Fig.~\ref{wings2}) contribute to the ponderomotive fluctuation force impinged by the light on the mirror. We marked with a tilde these complex amplitudes of the input and output waves for the sake of shortness. We neglected by these harmonic in the analysis above because the amplitude of the harmonics can be small. In what follow we take them into account and find the limitations they introduce for the proposed here measurement strategy. 
 
 Parasitic side bands provide additional terms to the expressions of the optical fields, for instance, \eqref{beta-a2} should be rewritten as  
  \na{
 \begin{align}
  \label{beta-a3}
  \beta_{-a}^\text{comb} &= \xi\left(\alpha_{-a} 
    + \frac{ \mathcal K  \tilde g_{a+}}{(\gamma_m-i\Omega)} 
        \cdot \frac{(\gamma -i\Omega)}{\sqrt{2\gamma}} -\right.\nonumber\\
     &\left.- \frac{\sqrt{\xi \mathcal K }}{\gamma_m-i\Omega}
            \left(\sqrt {2 \gamma_m} q_a - f_{s\,\phi}\right)\right),
 \end{align}
}
where noise term $\tilde g_{a_+}$, defined by \eqref{tildeg}, at condition \eqref{RSB} is approximately equal to 
\begin{align}
 \tilde g_{a+} &\simeq \frac{\sqrt \gamma}{2\omega_m}
        \left[\tilde a_{\phi+} - \tilde a_{\phi-}\right]
\end{align}
See details in Appendix~\ref{AppA} for details.

The back action created by the parasitic side bands limits sensitivity of the measurements. Instead of \eqref{Sf} we obtain a corrected expression for the single-sided power spectral density that can be presented as
\begin{align}
  S_{f}(\Omega) &= 2\gamma_m\big(2n_T+1\big) 
         + \frac{\big(\gamma_m^2+\Omega^2\big)}{|\mathcal K|}
         + \frac{|\mathcal K|(\gamma^2+ \Omega^2)}{4\omega_m^2}\ge \nonumber\\
         &\ge 2\gamma_m\big(2n_T+1\big) 
           + \frac{\sqrt{\gamma^2+ \Omega^2}}{2\omega_m}\cdot S_{SQL,f}  \label{Sf2}
 \end{align}
While the sensitivity still can be less than SQL at conditions \eqref{RSB}, the sensitivity becomes limited after the probe power reaches the optimal value found from Eq.~\eqref{Sf2}
\begin{equation}
    |\mathcal K|^2=4\omega_m^2 \frac{\gamma_m^2+\Omega^2}{\gamma^2+\Omega^2} \gg |{\mathcal K}_{SQL}|^2,
\end{equation}
where ${\mathcal K}_{SQL}$ corresponds the optimal power value needed to reach SQL in the system.

The impact of the parasitic harmonics can be reduced if one is able to measure $\tilde b_-(\Omega)= b_-(\omega_-+2 \omega_m+\Omega)$ as well as $\tilde b_+(\Omega) = b_+(\omega_- - 2 \omega_m+\Omega)$ independently on the other spectral components of the output light. The measurement can be performed if narrow band band pass filters are available. 

The scheme of such a measurement is illustrated by Fig.~\ref{wings2}. Measurement of optimally selected quadratures of $\tilde b_\pm$ allows partial reduction of the parasitic back action described by term $\sim \tilde g_{a+}$  in \eqref{beta-a3}. The reduction factor is $R\simeq \Omega/2\omega_m \ll 1$ (see details in Appendix~\ref{AppB}). 

It means that in case $\gamma_m=0$ and at conditions \eqref{RSB} the formula \eqref{Sf2} will have form
\begin{align}
  \label{Sf3}
  S_{f}(\Omega) &=  \frac{\Omega^2}{|\mathcal K|}
         + \frac{|\mathcal K|\gamma^2 \Omega^2}{16\omega_m^4}
         \ge  \frac{\gamma|\Omega|}{2\omega_m^2}\cdot S_{SQL,f}
\end{align}
The sensitivity \eqref{Sf3} is better than one defined by \eqref{Sf2} achieved for the case of not suppressed parasitic harmonics.

\section{Discussion and conclusion}

In the transducers shown in the schemes Fig.~\ref{TriCav} and Fig.~\ref{Cav2modes} the information on the mechanical {\em quadratures} transfers to the optical quadratures. The measurement of the difference of the optical amplitude quadratures is equivalent to the registration of the mechanical  amplitude quadrature, whereas the measurement of the sum of the optical phase quadratures  corresponds to the registration of mechanical  phase quadrature, as shown by Eq.~\eqref{quadIn}. This is a peculiar property of the parametric interaction. 

One of the main features of the proposed here measurement strategy is in the usage of the dichromatic probe field results in the two independent quantum outputs. It gives us a flexibility to measure back action separately and then subtract it completely from the measurement result. The subtraction can be made in a broad frequency band.

In contrast, in conventional variational measurements \cite{93a1VyMaJETP, 95a1VyZuPLA, 02a1KiLeMaThVyPRD} there is only one quantum output and the back action cannot be measured separately from the signal. Measurement of the linear combination of the amplitude and phase quadratures in that case allows partial subtraction of the back action. Only one quadrature of the output wave has to be measured to surpass SQL.

The scheme proposed here allows measurement either of a combination of sum and difference amplitude quadratures \eqref{beta-a2}  or sum and difference of phase  quadratures \eqref{betaMSIphi}. Generalization \eqref{quadphi} is also possible. These measurement lead to back action evasion in a broad frequency band. We expect that this technique will find a realization in other metrological configurations.

We propose to use filtration of output waves in order to depress back action due to parasitic side bands. Experimental realization of this filtration is not a simple task, but it is possible {\em in principle}.

 \acknowledgments
The research of SPV and AIN has been supported by the Russian Foundation for Basic Research (Grant No. 19-29-11003), the Interdisciplinary Scientific and Educational School of Moscow University ``Fundamental and Applied Space Research'' and from the TAPIR GIFT MSU Support of the California Institute of Technology. The reported here research performed by ABM was carried out at the Jet Propulsion Laboratory, California Institute of Technology, under a contract with the National Aeronautics and Space Administration (80NM0018D0004). This document has LIGO number ???.  

\appendix

\section{Derivation of the intracavity fields}
\label{IntrDeriv}
 
Here we provide details of calculation for intracavity fields, for example, see \cite{Walls2008}.

We begin with Hamiltonian \eqref{Halt}
\begin{align}  
  H  & = H_0 + H_\text{int}+H_T+H_\gamma+H_{T, \, m}+H_{\gamma_m}, \\
   \label{Ht}
      H_T & = \sum\limits_{k=0}^\infty \hslash \omega_k b_k^\dag b_k,\\
      \label{Hgamma}
      H_\gamma & =i \hslash \sqrt\frac{\gamma \Delta \omega}{\pi} \sum\limits_{k=0}^\infty \left[(c_+^\dag+c_-^\dag) b_k-(c_++c_-)b_k^\dag \right],\\
       \label{Htm}
      H_{T, \, m} & = \sum\limits_{k=0}^\infty \hslash \omega_k q_k^\dag q_k,\\
      \label{Hgammam}
      H_{\gamma_m} & =i \hslash \sqrt\frac{\gamma \Delta \omega}{\pi} \sum\limits_{k=0}^\infty \left[d^\dag q_k-dq_k^\dag \right].
  \end{align}
Here $H_T$ is the Hamiltonian of the environment presented as a bath of  oscillators described with frequencies $\omega_k=\omega_{k-1}+\Delta \omega$ and annihilation and creation operators $b_k$, $b_k^\dag$.  $H_\gamma$ is the Hamiltonian of coupling between the environment and the optical resonator, $\gamma$ is the coupling constant. Similarly $H_{T, \, m}$ is the Hamiltonian of the environment presented by a thermal bath of mechanical oscillators with frequencies $\omega_k=\omega_{k-1}+\Delta \omega$ and amplitudes described with annihilation and creation operators $q_k$, $q_k^\dag$.  $H_{\gamma_m}$ is the Hamiltonian of coupling between the environment and the mechanical oscillator, $2\gamma_m$ is the decay rate of the oscillator.
  
We write Heisenberg equations for operators $c_+$ and $b_k$:
\begin{subequations} 
  \begin{align}
i\hslash \dot c_+ & =[c_+,H]=\hslash\omega_+ c_+-i\hslash \eta c_- d+i \hslash \sqrt\frac{\gamma \Delta \omega}{\pi} \sum\limits_{k=0}^\infty b_k,\\
i \hslash \dot b_k & =[b_k,H]=\hslash \omega_k b_k-i \hslash \sqrt\frac{\gamma \Delta \omega}{\pi}  \left(c_++c_-\right)
  \end{align} \label{Heis1}
  \end{subequations}

We introduce slow amplitudes $c_\pm \rightarrow c_\pm e^{-i \omega_\pm t}$, $d \rightarrow d e^{-i(\omega_+-\omega_-)t}$, $b_k \rightarrow b_k e^{-i \omega_k t}$ and substitute them into \eqref{Heis1}
\begin{subequations} 
  \begin{align}
\dot c_+ & =-\eta c_- d+  \sqrt\frac{\gamma \Delta \omega}{\pi} \sum\limits_{k=0}^\infty b_k e^{-i(\omega_k-\omega_+)t}, \label{Heisc}\\
  \dot b_k & = -\sqrt\frac{\gamma \Delta \omega}{\pi} \left(c_+e^{-i(\omega_+-\omega_k)t}+c_- e^{-i(\omega_--\omega_k)t}\right) \label{Heisb}
  \end{align}
  \end{subequations}

Using initial condition $b_k(t=0)=b_k(0)$ to integrate \eqref{Heisb} we obtain
\begin{align}
b_k(t) & =b_k(0)-\int\limits_0^t\sqrt\frac{\gamma \Delta \omega}{\pi}  c_+(s)e^{-i(\omega_+-\omega_k)s}ds-\\
& -\int\limits_0^t\sqrt\frac{\gamma \Delta \omega}{\pi}c_-(s)e^{-i(\omega_--\omega_k)s}ds \label{binit}
\end{align} 
Using final condition $b_k(t=\infty)=b_k(\infty)$ to integrate \eqref{Heisb} we derive
\begin{align}
b_k(t) & =b_k(\infty)+\int\limits_t^\infty\sqrt\frac{\gamma \Delta \omega}{\pi}  c_+(s)e^{-i(\omega_+-\omega_k)s}ds-\\
& +\int\limits_t^\infty\sqrt\frac{\gamma \Delta \omega}{\pi}c_-(s)e^{-i(\omega_--\omega_k)s}ds \label{bfin}
\end{align} 
To get the input-output relation we substitute initial condition \eqref{binit} into \eqref{Heisc}
\begin{subequations} 
  \begin{align}
\dot c_+ & =-\eta c_- d+   \sum\limits_{k=0}^\infty \sqrt\frac{\gamma \Delta \omega}{\pi} b_k(0)e^{-i(\omega_k-\omega_+)t}-\\
&-\sum\limits_{k=0}^\infty\int\limits_0^t\frac{\gamma \Delta \omega}{\pi}  c_+(s) 
	 e^{-i(\omega_k-\omega_+)(t-s)}ds -\\
&-\left(\sum\limits_{k=0}^\infty\int\limits_0^t\frac{\gamma \Delta \omega}{\pi}  c_-(s) e^{-i(\omega_k-\omega_-)(t-s)}ds \right)e^{i(\omega_+-\omega_-)t}\nonumber
  \end{align} \label{Heis2}
  \end{subequations}
and omit the last term proportional to $e^{i(\omega_+-\omega_-)t}$ as fast oscillating, and define the input field
  \begin{equation}
a_+(t) = \sum\limits_{k=0}^\infty \sqrt\frac{ \Delta \omega}{2\pi} b_k(0)e^{-i(\omega_k-\omega_+)t} \label{aplus}
\end{equation}
To calculate the remaining sum in \eqref{Heis2} we  assume the limit $\Delta \omega \rightarrow 0$ and  replace the sum by the integral using rule
\begin{subequations} 
  \begin{align} 
\Delta \omega &\sum\limits_{k=0}^\infty \rightarrow \int\limits_0^\infty d \omega_k\\
&\sum\limits_{k=0}^\infty\int\limits_0^t\frac{\gamma \Delta \omega}{\pi}  c_+(s) e^{-i(\omega_k-\omega_+)(t-s)}ds  \rightarrow\\ \nonumber
\rightarrow & \int\limits_0^\infty\int\limits_0^t 2\gamma  c_+(s) e^{-i(\omega_k-\omega_+)(t-s)}ds \frac{d \omega_k}{2 \pi}=\\ \nonumber
= &  \int\limits_{-\omega_+}^\infty\int\limits_0^t 2\gamma  c_+(s) e^{-i \omega(t-s)}ds \frac{d \omega}{2 \pi}\approx\\ \nonumber
\approx &  \int\limits_{-\infty}^\infty\int\limits_0^t 2\gamma  c_+(s) e^{-i \omega(t-s)}ds \frac{d \omega}{2 \pi}=\\
= & \int\limits_0^t 2\gamma  c_+(s) \delta(t-s) ds = \frac{2 \gamma c_+(t)}{2}=\gamma c_+(t)
  \end{align} \label{gammac}
  \end{subequations}

Substituting \eqref{aplus} and \eqref{gammac} into \eqref{Heis2} we obtain
   \begin{align}
\dot c_+ & =-\eta c_- d+  \sqrt{2\gamma}a_+-\gamma c_+\\
\dot c_+ & + \gamma c_++\eta c_- d=  \sqrt{2\gamma}a_+. \label{cplus}
\end{align}

By analogue we derive the equation for input field $a_-$ and present it in the similar form
  \begin{equation}
a_-(t) = \sum\limits_{k=0}^\infty \sqrt\frac{ \Delta \omega}{2\pi} b_k(0)e^{-i(\omega_k-\omega_-)t}
\end{equation}
It leads to the equation for the intracavity field $c_-$
   \begin{align}
\dot c_-+ \gamma c_--\eta c_+ d^\dag=  \sqrt{2\gamma}a_-. \label{cminus}
\end{align}

Similar equation can be derived for the amplitude $q(t)$ of the mechanical oscillator
  \begin{equation}
q(t)=  \sum\limits_{k=0}^\infty \sqrt\frac{ \Delta \omega}{2\pi} b_{m, \,k}(0)e^{-i(\omega_k-\omega_+)t}
\end{equation}
and get Langevine  equation for mechanical oscillator's quadrature $d$
   \begin{align}
\dot d+ \gamma_m d-\eta^* c_+ c_-^\dag=  \sqrt{2\gamma_m}q.
\end{align}
To get the output relation we substitute \eqref{bfin} into \eqref{Heisc} and define the output fields 
  \begin{align}
b_+(t) = -\sum\limits_{k=0}^\infty \sqrt\frac{ \Delta \omega}{2\pi} b_k(\infty)e^{-i(\omega_k-\omega_+)t} \\
b_-(t) = -\sum\limits_{k=0}^\infty \sqrt\frac{ \Delta \omega}{2\pi} b_k(\infty)e^{-i(\omega_k-\omega_+)t}.
\end{align}
It leads to 
   \begin{align} 
\dot c_+ & - \gamma c_++\eta c_- d=  \sqrt{2\gamma}b_+ \label{bplus}\\
\dot c_- & - \gamma c_-+\eta^* c_+ d^\dag=  \sqrt{2\gamma}b_- \label{bminus}
\end{align}
Utilizing pairs of equations \eqref{cplus} and \eqref{bplus}  as well as \eqref{cminus} and \eqref{bminus} we obtain the final expression for the input-output relations
\begin{align}
b_+=-a_++\sqrt{2\gamma} c_+\\
b_-=-a_-+\sqrt{2\gamma} c_-
\end{align}

Let us to derive the commutation relations for the Fourier amplitudes of the operators. We introduce Fourier transform of field $a_+(t)$ using \eqref{aplus}
 \begin{subequations}
 \begin{align}
a_+(\Omega) = \int\limits_{-\infty}^{\infty} \sum\limits_{k=0}^\infty \sqrt\frac{ \Delta \omega}{2\pi} b_k(0)e^{-i(\omega_k-\omega_+-\Omega)t}dt=\\
= \sum\limits_{k=0}^\infty \sqrt { 2 \pi \Delta \omega}  b_k(0)  \delta(\Omega-\omega_k+\omega_+)
 \end{align}
 \end{subequations}
 This allows us to find the commutators \eqref{comm1}: 
  \begin{subequations}
 \begin{align} \nonumber
 [a_+(\Omega)& ,a_+^\dag(\Omega') ] 
   =\sum\limits_{k=0}^\infty  2 \pi \Delta \omega   [b_k(0),b_k^\dag(0)] \times\\ \nonumber
&\times \delta(\Omega-\omega_k+\omega_+)  \delta(\Omega'-\omega_k+\omega_+) \rightarrow\\ \nonumber
&\rightarrow \int\limits_{-\infty}^\infty 2 \pi [b(0),b^\dag(0)]  \delta(\Omega-\omega)  \delta(\Omega'-\omega) d\omega=\\
&=2\pi \delta(\Omega-\Omega'),
 \end{align}
 \end{subequations}
 and the correlators \eqref{corr1}
   \begin{subequations}
 \begin{align} \nonumber
\langle a_+(\Omega)&,a_+^\dag(\Omega') \rangle
 = \sum\limits_{k=0}^\infty  2 \pi \Delta \omega   \langle b_k(0),b_k^\dag(0) \rangle \times\\ \nonumber
&\times \delta(\Omega-\omega_k+\omega_+)  \delta(\Omega'-\omega_k+\omega_+) \rightarrow\\ \nonumber
&\rightarrow \int\limits_{-\infty}^\infty 2 \pi \langle b(0),b^\dag(0) \rangle \delta(\Omega-\omega)  \delta(\Omega'-\omega) d\omega=\\
&=2\pi \delta(\Omega-\Omega')
 \end{align}
 \end{subequations}
Similar expressions can be derived for commutators and correlators of the optical $a_-$ and mechanical $q$ quantum amplitudes.

\section{Suppression of the resonant ponderomotive excitation}\label{AppD}

In this section we discuss possibilities of suppression of resonant ponderomotive excitation of the mechanical oscillations of the system.

We assume that there exist modes $c_+,\ e_+$ characterized with orthogonal polarizations and the same geometrical path. These modes are characterized with eigen frequency $\omega_+$. Similarly, modes $c_-,\ e_-$ are characterized with orthogonal polarizations and the same eigen frequency $\omega_-$. In this configuration
\eqref{Halt}:
 \begin{subequations}
   \label{Halt2}
  \begin{align}
  H & = \tilde H_0 + \tilde H_\text{int} + H_\gamma,\\
  \tilde H_0 &=\hslash \omega_+c_+^\dag c_+  + \hslash \omega_-c_-^\dag c_-\\
  \label{H02}
      &\quad    + \hslash \omega_+e_+^\dag e_+  + \hslash \omega_-e_-^\dag e_-
         +\hslash \omega_m d^\dag d,\\
      \label{Hint2a}
     \tilde H_\text{int} & = \frac{\hslash }{i}
      \left(\eta c_+^\dag c_-d - \eta^* c_+ c_-^\dag d^\dag \right)+\\
     \label{Hint2b}
      &\qquad + \frac{\hslash }{i}
      \left(\eta_e e_+^\dag e_-d - \eta_e^* e_+ e_-^\dag d^\dag \right)  
  \end{align}
 \end{subequations}
For the mean amplitude $D$ we derive
\begin{align}
 \label{D2}
 \gamma_m D &=\eta^* C_+ C_-^* + \eta_e E_+E_-^*
\end{align}
Optimizing pumps of additional modes in a proper way one can make right part in \eqref{D2} to be zero.

The technique of suppression of the resonant ponderomotive excitation of the mechanical oscillations should not introduce undesirable fluctuations that cannot be removed by the measurement procedure. There at least two possibilities:
\begin{itemize}
 \item Coupling constants are equal $\eta= \eta_e$ for both parts \eqref{Hint2a} and \eqref{Hint2b} of $\tilde H_\text{int}$. The additional back action originated from modes $e_\pm$ can be evaded if we measure outputs of the four modes simultaneously. 
 \item Coupling constant $\eta_e$ is small ($\eta_e\ll \eta$). It can be realized by engineering coating of mirror $M$ on Fig.~\ref{Cav2modes}. The larger pumps $E_\pm$ can compensate regular force in \eqref{D2}, without introducing significant back action noise. 
\end{itemize}

\section{Account of parasitic side bands}\label{AppA}

In this section we provide details of calculations taking into account the parasitic optical harmonics in the system.

We keep in mind that fluctuation fields in ``$+$'' and ``$-$'' cavities are not correlated with each other. Let us generalize interaction Hamiltonian \eqref{Hint}, assuming coupling constant $\eta$ to be real:
\begin{subequations}
 \begin{align}
  \label{Hint3a1}
  &\tilde H_{int}
   \simeq \frac{\hslash \eta}{i}\left\{ 
      \left[C_+^*c_- + C_-c_+^\dag\right]d - \left[C_-^*c_+ + C_+c_-^\dag \right]d^\dag +\right.\\
    \label{Hint3b1}
      & \left.+ \left[C_+ ^*\tilde c_- + C_-\tilde c_+^\dag\right]d^\dag
        -\left[C_+ \tilde c_-^\dag + C_-^*\tilde c_+\right]d
   \right \}
 \end{align}
Here terms \eqref{Hint3a1} correspond to conventional opto-mechanical coupling including back action, whereas terms \eqref{Hint3b1} corresponds to parasitic back action due to the high frequency side bands $\tilde c_\pm$. 
\end{subequations}

The additional  terms appear in the set of Hamilton equations \eqref{dd} for the mechanical operators
\begin{subequations}
 \label{ddb}
 \begin{align} 
  (\gamma_m - i\Omega)  d(\Omega) &=\eta\left[C_+ c_{-}^\dag(-\Omega) + C_-^* c_+(\Omega)\right]-\\
        \label{ddbtilde}
        & - \eta\left[ C_+^* \tilde c_{-}(\Omega)- C_- \tilde c_{+}^\dag(-\Omega)\right] +\\
        &+\sqrt{2\gamma_m} q(\Omega) + i f_s(\Omega),
  \end{align}
\end{subequations}
where terms \eqref{ddbtilde} describe the contributions due to the parasitic back action due to \eqref{Hint3b1}.

The operators $c_\pm$ obey \eqref{set1}. For the operators of the parasitic sidebands inside the cavity we derive
\begin{subequations}
 \label{tildec}
 \begin{align} 
  \tilde c_+&(\Omega)\big(\gamma +i2\omega_m -i\Omega\big) = \sqrt{2\gamma}\, \tilde a_+(\Omega)-\\ 
       & -\eta C_-d^\dag(-\Omega)  ,\quad \tilde a_+(\Omega) = a_+(-2\omega_m + \Omega),\\
  \tilde c_-&(\Omega) \big(\gamma - 2i\omega_m -i\Omega\big) 
      = \sqrt{2\gamma}\, \tilde a_-(\Omega) +\\
      & +  \eta C_+d (\Omega),\quad   \tilde a_-(\Omega) = a_-(2\omega_m + \Omega).
 \end{align}
\end{subequations}
For {\em output} amplitudes of parasitic sidebands we find
\begin{subequations}
 \label{tildeb}
 \begin{align}
  \tilde b_+(\Omega) &= \frac{\gamma -2i\omega_m  +i\Omega}{\gamma +2i\omega_m -i\Omega}\, \tilde a_+(\Omega)  
        -\frac{\sqrt{2\gamma}\,\eta C_-d^\dag(-\Omega)}{\gamma +2i\omega_m -i\Omega} \, ,\\
  \tilde b_-(\Omega)
      &= \frac{\gamma +2i\omega_m  +i\Omega}{\gamma -2i\omega_m -i\Omega}\,  \tilde a_-(\Omega) 
       + \frac{\sqrt{2\gamma}\, \eta C_+d (\Omega)}{\gamma -2i\omega_m -i\Omega}\,,
 \end{align}
\end{subequations}

\subsection{Quadratures}

Let us consider the case of resonance probe light \eqref{real}, find quadratures for parasitic optical harmonics and  rewrite the expression \eqref{ddb} for the mechanical quadratures as
\begin{subequations}
 \label{ddquad}
 \begin{align}
  \tilde c_{a\pm} &=\frac{\tilde c_\pm(\Omega) + \tilde c_\pm^\dag(-\Omega)}{\sqrt 2},\quad
     \tilde c_{\phi\pm} =\frac{\tilde c_\pm(\Omega) - \tilde c_\pm^\dag(-\Omega)}{i\sqrt 2}\\
  (\gamma_m -& i\Omega)  d_a=\eta C\left[ c_{a-} + c_{a+}\right]-\\
         &-  \eta C\left[\tilde c_{a-}- \tilde c_{a+}\right] 
         +\sqrt{2\gamma_m}\, q_a  - f_{\phi s}, \\
 (\gamma_m -& i\Omega) d_\phi = \eta C\left[- c_{\phi-} +  c_{\phi+}\right]-\\
        & -  \eta C\left[ \tilde c_{\phi-} +  \tilde c_{\phi+}\right] 
		  +\sqrt{2\gamma_m}\, q_\phi + f_{as} 
  \end{align}
\end{subequations}
Substituting \eqref{ddquad} into (\ref{c+}, \ref{c-}) we get
\begin{subequations}
 \begin{align}
    c_{a+} &=  \frac{\sqrt {2 \gamma} \,  a_{a+}}{\gamma -i\Omega} 
    - \frac{\eta^2 C^2\left[ c_{a-} +  c_{a+}
           -\tilde c_{a-}-  \tilde c_{a+}\right]}{(\gamma_m-i\Omega)(\gamma -i\Omega)} -\nonumber\\
       &\quad  -\eta C\cdot\frac{\sqrt{2\gamma_m}\, q_a  -f_{\phi s}}{(\gamma_m-i\Omega)(\gamma -i\Omega)}\\
    c_{a-} &=  \frac{\sqrt {2 \gamma} \,  a_{a-}}{\gamma -i\Omega} 
    + \frac{\eta^2 C^2\left[ c_{a-} + c_{a+}
           - \tilde c_{a-}- \tilde c_{a+}\right]}{(\gamma_m-i\Omega)(\gamma -i\Omega)} -\nonumber\\
       &\quad +\eta C\cdot\frac{\sqrt{2\gamma_m}\, q_a  -f_{\phi s}}{(\gamma_m-i\Omega)(\gamma -i\Omega)}\\
   c_{\phi+} &=  \frac{\sqrt {2 \gamma} \,  a_{\phi+}}{\gamma -i\Omega} 
    - \frac{\eta^2 C^2\left[- c_{\phi-} +  c_{\phi+}
           - \tilde c_{\phi-} +  \tilde c_{\phi+}\right]}{(\gamma_m-i\Omega)(\gamma -i\Omega)} - \nonumber\\
       &\quad  -\eta C\cdot\frac{\sqrt{2\gamma_m}\, q_\phi  +f_{a s}}{(\gamma_m-i\Omega)(\gamma -i\Omega)}\\
   c_{\phi-} &=  \frac{\sqrt {2 \gamma} \,  a_{\phi-}}{\gamma -i\Omega} 
    - \frac{\eta^2 C^2\left[- c_{\phi-} + c_{\phi+}
         - \tilde c_{\phi-} +  \tilde c_{\phi+}\right]}{(\gamma_m-i\Omega)(\gamma -i\Omega)} -\nonumber\\
       &\quad  -\eta C\cdot\frac{\sqrt{2\gamma_m}\, q_\phi  +f_{a s}}{(\gamma_m-i\Omega)(\gamma -i\Omega)}
 \end{align}
\end{subequations}

We introduce sum and difference quadratures for the parasitic sidebands, similarly to (\ref{gDef}, \ref{alphaDef}, \ref{betaDef}):
\begin{subequations}
 \begin{align}
 \label{gDef2}
  \tilde g_{a\pm}& = \frac{\tilde c_{+a}\pm \tilde c_{-a}}{\sqrt 2},\quad 
  \tilde g_{\phi\pm} = \frac{\tilde c_{+\phi}\pm \tilde c_{-\phi}}{\sqrt 2},\\
 \label{alphaDef2}
    \tilde \alpha_{a\pm}& = \frac{\tilde a_{+a}\pm \tilde a_{-a}}{\sqrt 2}\,,\quad
    \tilde \alpha_{\phi\pm}= \frac{\tilde a_{+\phi}\pm \tilde a_{-\phi}}{\sqrt 2},\\
 \label{betaDef2}
    \tilde \beta_{a\pm}&= \frac{\tilde b_{+a}\pm \tilde b_{-a}}{\sqrt 2}\,,\quad
    \tilde \beta_{\phi\pm}= \frac{\tilde b_{+\phi}\pm \tilde b_{-\phi}}{\sqrt 2}
\end{align}
\end{subequations}
and arrive to the expressions for the sum and difference quadratures
\begin{subequations}
 \label{cdiffsum}
 \begin{align}
   g_{a+} &  =  \frac{\sqrt{2 \gamma} \, \alpha_{a+}}{\gamma -i\Omega}\, , \\
\label{ga-b}
   g_{a-}& =  \frac{\sqrt {2 \gamma} \,  \alpha_{a-}}{\gamma -i\Omega} 
    - \frac{ 2\,\eta^2 C^2 \left[g_{a+} -\tilde g_{a+}\right]}{(\gamma_m-i\Omega)(\gamma -i\Omega)} -\\
      &\quad   -\sqrt2\,\eta C\cdot
        \frac{\sqrt{2\gamma_m}\, q_a - f_{\phi s}}{(\gamma_m-i\Omega)(\gamma -i\Omega)},\\
   g_{\phi-} & =  \frac{\sqrt{2 \gamma} \, \alpha_{\phi-}}{\gamma -i\Omega},\\
   g_{\phi+} & =  \frac{\sqrt {2 \gamma} \,  \alpha_{\phi+}}{\gamma -i\Omega} 
    - \frac{ 2\eta^2 C^2\left[g_{\phi-}+ \tilde g_{\phi-}\right]}{(\gamma_m-i\Omega)(\gamma -i\Omega)} -\\
      &\quad -\sqrt 2\, \eta C\cdot
        \frac{\sqrt{2\gamma_m}\, q_\phi  +f_{a s}}{(\gamma_m-i\Omega)(\gamma -i\Omega)}.
 \end{align}
\end{subequations}
At the next step we evaluate the sum and difference quadratures of side bands using \eqref{tildec}
\begin{subequations}
 \label{tildeg}
 \begin{align}
  \tilde g_{a+} &= \frac{\sqrt{2\gamma}}{ 2}
     \left(\frac{\tilde a_+(\Omega) +\tilde a_-^\dag(-\Omega)}{
            \big(\gamma +2i\omega_m -i\Omega\big)} 
       + \frac{\tilde a_+^\dag(-\Omega) +\tilde a_-(\Omega)}{
        \big(\gamma - 2i\omega_m -i\Omega\big)}
    \right),\\
  \tilde g_{\phi-} &=\frac{\sqrt{2\gamma}}{i\, 2}
     \left(\frac{\tilde a_+(\Omega) +\tilde a_-^\dag(-\Omega)}{ 
            \big(\gamma + 2 i\omega_m -i\Omega\big)} 
       - \frac{\tilde a_+^\dag(-\Omega) +\tilde a_-(\Omega)}{
        \big(\gamma - 2i\omega_m -i\Omega\big)}
    \right).
 \end{align}
The combinations above do not contain any information on displacement of the mechanical oscillator.
\end{subequations}

For output sum and difference quadratures we obtain taking advantage of the Eq. \eqref{output}
\begin{subequations}
 \begin{align}
  \label{bdiffsum1}
  \beta_{a-}& =  \frac{ \gamma +i\Omega} {\gamma -i\Omega} \, \,  \alpha_{a-}
    - \sqrt{2\gamma}\frac{ 2\eta^2 C^2 \left[ g_{a+} 
     -  \tilde g_{a+}\right]}{(\gamma_m-i\Omega)(\gamma -i\Omega)} -\\
    &\quad - 2\sqrt{\gamma}\,\eta C\cdot
        \frac{\sqrt{2\gamma_m}\, q_a -\,f_{\phi s}}{(\gamma_m-i\Omega)(\gamma -i\Omega)}\\
  \beta_{a+} & =  \frac{ \gamma+i\Omega }{\gamma -i\Omega}\, \alpha_{a+}\,,     \\
\label{bdiffsum2}
  \beta_{\phi+} & =  \frac{\gamma+i\Omega }{\gamma -i\Omega} \,\alpha_{\phi+}
    - \sqrt{2\gamma}\frac{ 2\eta^2 C^2\left[ g_{\phi-}  + \tilde g_{\phi-} \right]}{
	(\gamma_m-i\Omega)(\gamma -i\Omega)} -\nonumber\\
     &\quad -2\sqrt{ \gamma}\ \, \eta C\cdot
        \frac{\sqrt{2\gamma_m}\, q_\phi  +f_{a s}}{(\gamma_m-i\Omega)(\gamma -i\Omega)}\\
  g_{\phi-} &  =  \frac{ \gamma+ i\Omega}{\gamma -i\Omega}\,  \alpha_{\phi-}.
 \end{align}
 \end{subequations}
Finally, we rewrite $\beta_{a+}$ after compensation of the main back action term as \eqref{beta-a3}.

We see that one can subtract completely the main ($\sim \alpha_{a+}$) term of the back action. The contribution of the parasitic harmonic ($\sim \tilde g_{a+}$) into back action limit the sensitivity of the measurement in this case. In the following section we discuss a possibility of reduction of their impact.

\section{Reduction of the residual back action}\label{AppB}

Using \eqref{tildeg} we find
\begin{subequations}
 \begin{align}
  \tilde \beta_{a+} &=\frac{1}{2}
     \left\{\frac{\gamma -2i\omega_m +i\Omega}{
            \gamma +2i\omega_m -i\Omega} \cdot
            \big( \tilde a_+(\Omega) +\tilde a_-^\dag(-\Omega)\big)\right.\\
     & \left.  + \frac{\gamma + 2i\omega_m +i\Omega}{
         \gamma - 2i\omega_m -i\Omega}\cdot
         \big(\tilde a_+^\dag(-\Omega) +\tilde a_-(\Omega)\big)
      \right\},\\
  \tilde \beta_{\phi-} &=\frac{1}{2i}
     \left\{\frac{\gamma -2i\omega_m +i\Omega}{
            \gamma +2i\omega_m -i\Omega} 
            \big( \tilde a_+(\Omega) +\tilde a_-^\dag(-\Omega)\big)\right.\\
      &\left. - \frac{\gamma + 2i\omega_m -i\Omega}{
         \gamma - 2i\omega_m +i\Omega}\cdot
         \big(\tilde a_+^\dag(-\Omega) +\tilde a_-(\Omega)\big)
      \right\}.
 \end{align}
\end{subequations}
In order to compensate ``tilde'' terms we have to measure a combination of arbitrary quadratures defined by phases $\varphi_\pm$
 \begin{subequations}
 \begin{align}
  \tilde b_{\varphi+} &= \frac{ \tilde b_{+}(\Omega) e^{i\varphi_+}
      + \tilde b_{+}^\dag (\Omega) e^{-i\varphi_+}}{\sqrt 2},\\
  \tilde b_{\varphi-} & = \frac{ \tilde b_{-}(\Omega) e^{i\varphi_-}
      + \tilde b_{-}^\dag (\Omega) e^{-i\varphi_-}}{\sqrt 2}\,.
 \end{align}
In order to remove parasitic terms we have to define the phases  as follows
  \begin{align}
    & \varphi_+=-\varphi_- =\varphi\,\quad \Rightarrow\quad \\
   \frac{\tilde b_{\varphi+} + \tilde b_{\varphi-}}{\sqrt 2}  
     &= \frac{\gamma -i2\omega_m  +i\Omega}{2(\gamma +i2\omega_m -i\Omega)}
	\left( \tilde a_+(\Omega)+ \tilde a_-^\dag(-\Omega)\right)e^{i\varphi}
	\nonumber\\
    & + \frac{\gamma +i2\omega_m  +i\Omega}{2(\gamma -i2\omega_m -i\Omega)}
      \left(  \tilde a_-(\Omega) + \tilde a_+^\dag(-\Omega) \right) e^{-i\varphi}\nonumber
  \end{align}

We get for $\varphi$ an approximate expression
 \begin{align}
  \big(\gamma -i2\omega_m  +i\Omega\big)& e^{i\varphi}\simeq \text{const}, \\ 
  \Rightarrow\quad &
    e^{i\varphi} = \sqrt\frac{\gamma+2i\omega_m}{\gamma-2i\omega_m}
 \end{align}
\end{subequations}
Performing the same procedure we developed for the main harmonics of the probe light we find that the impact of the parasitic harmonics can be reduced by $R\simeq \Omega/2\omega_m$ (assuming validity of conditions \eqref{RSB}). After the compensation we obtain 
\begin{align}
 \label{betaa-2}
  \beta_{a-}& =  \frac{\xi\sqrt{\mathcal K}}{(\gamma_m-i\Omega)} \left\{  
        \frac {(\gamma_m-i\Omega)}{ \sqrt{\mathcal K} }\, \alpha_{a-}
        -\sqrt{\mathcal K}\, \alpha_{a+}
       \right.\\
    + \sqrt{\mathcal K} &\left.   \frac{(\gamma -i\Omega)}{\sqrt{2\gamma}} \tilde g_{a+} R 
    - \sqrt{\xi^{-1}} \left[
        \sqrt{2\gamma_m}\, q_a - \,f_{\phi s}\right]
        \right\}.\nonumber
\end{align}
instead of \eqref{beta-a3}.
The first term in braces $\sim \alpha_{a-} $ results from the quantum measurement noise and the second term $ \sim \alpha_{a+}$ describes back action that can be removed from the measurement results. The back action term due to the parasitic harmonics can be reduced. Optimization of contributions of these terms defines the ultimate sensitivity of the measurement technique that is better than the SQL.


\begin{thebibliography}{10}

\bibitem{aLIGO2013}
LVC-Collaboration, ``{Prospects for Observing and Localizing Gravitational-Wave
  Transients with Advanced LIGO and Advanced Virgo},'' {\em arXiv},
  vol.~1304.0670, 2013.

\bibitem{aLIGO2015}
{J. Aasi {\em et al} (LIGO Scientific Collaboration)} {\em et~al.}, ``{Advanced
  LIGO},'' {\em Classical and Quantum Gravity}, vol.~32, p.~074001, 2015.

\bibitem{MartynovPRD16}
D.~Martynov {\em et~al.}, ``{Sensitivity of the Advanced LIGO detectors at the
  beginning of gravitational wave astronomy},'' {\em Physical Review D},
  vol.~93, p.~112004, 2016.

\bibitem{AserneseCQG15}
F.~Asernese {\em et~al.}, ``{Advanced Virgo: a 2nd generation interferometric
  gravitational wave detector},'' {\em Classical and Quantum Gravity}, vol.~32,
  p.~024001, 2015.

\bibitem{DooleyCQG16}
K.~L. Dooley, J.~R. Leong, T.~Adams, C.~Affeldt, A.~Bisht, C.~Bogan,
  J.~Degallaix, C.~Graf, S.~Hild, and J.~Hough, ``{GEO 600 and the GEO-HF
  upgrade program: successes and challenges},'' {\em Classical and Quantum
  Gravity}, vol.~33, p.~075009, 2016.

\bibitem{AsoPRD13}
Y.~Aso, Y.~Michimura, K.~Somiya, M.~Ando, O.~Miyakawa, T.~Sekiguchi, and
  D.~Tats, ``{Interferometer design of the KAGRA gravitational wave
  detector},'' {\em Physical Review D}, vol.~88, p.~043007, 2013.

\bibitem{ForstnerPRL2012}
{S. Forstner and S. Prams and J. Knittel and E.D. van Ooijen and J.D. Swaim and
  G.I. Harris and A. Szorkovszky and W.P. Bowen and H. Rubinsztein-Dunlop},
  ``{Cavity Optomechanical Magnetometer},'' {\em Physical Review Letters},
  vol.~108, p.~120801, 2012.

\bibitem{LiOptica2018}
B.-B. Li, J.~Bílek, U.~Hoff, L.~Madsen, S.~Forstner, V.~Prakash,
  C.~Schafermeier, T.~Gehring, W.~Bowen, and U.~Andersen, ``{Quantum enhanced
  optomechanical magnetometry},'' {\em Optica}, vol.~5, p.~850, 2018.

\bibitem{WuPRX2014}
{M. Wu}, {A.C. Hryciw}, {C. Healey}, {D.P. Lake}, {H. Jayakumar}, {M.R.
  Freeman}, {J.P. Davis}, and {P.E. Barclay}, ``{Dissipative and dispersive
  optomechanics in a nanocavity torque sensor},'' {\em Physical Review X},
  vol.~4, p.~021052, 2014.

\bibitem{KimNC2016}
P.~H. Kim, B.~D. Hauer, C.~Doolin, F.~Souris, and J.~P. Davis, ``{Approaching
  the standard quantum limit of mechanical torque sensing},'' {\em Nature
  Communications}, vol.~7, p.~13165, 2016.

\bibitem{AhnNT2020}
J.~Ahn, Z.~Xu, J.~Bang, P.~Ju, X.~Gao, and T.~Li, ``{Ultrasensitive torque
  detection with an optically levitated nanorotor},'' {\em Nature
  Nanotechnoligy}, vol.~15, p.~89–93, 2020.

\bibitem{Braginsky68}
{V.B. Braginsky}, ``{Classic and quantum limits for detection of weak force on
  acting on macroscopic oscillator},'' {\em Sov. Phys. JETP}, vol.~26,
  p.~831–834, 1968.

\bibitem{BrKh92}
{V.B. Braginsky} and {F.Ya. Khalili}, {\em {Quantum Measurement}}.
\newblock Cambridge University Press, Cambridge, 1992.

\bibitem{90BrKhPLA}
{V.B. Braginsky} and {F.Ya. Khalili.}, ``{Gravitational wave antenna with QND
  speed meter},'' {\em Physics Letters A}, vol.~147, p.~251–256, 1990.

\bibitem{00a1BrGoKhThPRD}
{V.B. Braginsky}, {M.L. Gorodetsky}, {F.Y. Khalili}, and {K.S. Thorne},
  ``{Dual-resonator speed meter for a free test mass},'' {\em Physical Review
  D}, vol.~61, p.~044002, 2000.

\bibitem{99a1BrKhPLA}
{V.B. Braginsky} and {F.Ya. Khalili}, ``{Low noise rigidity in quantum
  measurements},'' {\em Phys. Lett. A}, vol.~257, p.~241, 1999.

\bibitem{01a1KhPLA}
{F.Ya. Khalili}, ``{Frequency-dependent rigidity in large-scale interferometric
  gravitational-wave detectors},'' {\em Physics Letters A}, vol.~288,
  p.~251–256, 2001.

\bibitem{LigoNatPh11}
{The LIGO Scientific collaboration}, ``{A gravitational wave observatory
  operating beyond the quantum shot-noise limit},'' {\em Nature Physics},
  vol.~73, p.~962–965, 2011.

\bibitem{LigoNatPhot13}
{LIGO Scientific Collaboration and Virgo Collaboration}, ``{Enhanced
  sensitivity of the LIGO gravitational wave detector by using squeezed states
  of light},'' {\em Nature Photonics}, vol.~7, p.~613–619, 2013.

\bibitem{TsePRL19}
V.~Tse {\em et~al.}, ``{Quantum-Enhanced Advanced LIGO Detectors in the Era of
  Gravitational-Wave Astronomy},'' {\em Physical Review Letters}, vol.~123,
  p.~231107, 2019.

\bibitem{AsernesePRL19}
F.~Asernese, {et al}, and {(Virgo Collaboration)}, ``{Increasing the
  Astrophysical Reach of the Advanced Virgo Detector via the Application of
  Squeezed Vacuum States of Light},'' {\em Physical Review Letters}, vol.~123,
  p.~231108, 2019.

\bibitem{YapNatPhot20}
M.~Yap, J.~Cripe, G.~Mansell, {\em et~al.}, ``{Broadband reduction of quantum
  radiation pressure noise via squeezed light injection},'' {\em Nature
  Photonics}, vol.~14, p.~19–23, 2020.

\bibitem{YuNature20}
H.~Yu, L.~McCuller, M.~Tse, {\em et~al.}, ``{Quantum correlations between light
  and the kilogram-mass mirrors of LIGO},'' {\em Nature}, vol.~583, p.~43–27,
  2020.

\bibitem{CripeNat19}
J.~Cripe, N.~Aggarwal, R.~Lanza, {\em et~al.}, ``{Measurement of quantum back
  action in the audio band at room temperature},'' {\em Nature}, vol.~568,
  p.~364–367, 2019.

\bibitem{93a1VyMaJETP}
{S.P. Vyatchanin} and {A.B. Matsko}, ``{Quantum limit of force measurement},''
  {\em Sov.Phys – JETP}, vol.~77, p.~218–221, 1993.

\bibitem{95a1VyZuPLA}
S.~Vyatchanin and E.~Zubova, ``{Quantum variation measurement of force},'' {\em
  Physics Letters A}, vol.~201, p.~269–274, 1995.

\bibitem{02a1KiLeMaThVyPRD}
{H.J. Kimble}, Y.~Levin, {A.B. Matsko}, {K.S. Thorne}, and {S.P. Vyatchanin},
  ``{Conversion of conventional gravitational-wave interferometers into QND
  interferometers by modifying input and/or output optics},'' {\em Phys. Rev.
  D}, vol.~65, p.~022002, 2001.

\bibitem{TsangPRL2010}
M.~Tsang and C.~Caves, ``{Coherent Quantum-Noise Cancellation for
  Optomechanical Sensors},'' {\em Phys. Rev. Lett.}, vol.~105, p.~123601, 2010.

\bibitem{PolzikAdPh2014}
E.~Polzik and K.~Hammerer, ``{Trajectories without quantum uncertainties},''
  {\em Annalen de Physik}, vol.~527, p.~A15–A20, 2014.

\bibitem{MollerNature2017}
C.~Moller, R.~Thomas, G.~Vasilakis, E.~Zeuthen, Y.~Tsaturyan, M.~Balabas,
  K.~Jensen, A.~Schliesser, K.~Hammerer, and E.~Polzik, ``{Quantum
  back-action-evading measurement of motion in a negative mass reference
  frame},'' {\em Nature}, vol.~547, p.~191–195, 2017.

\bibitem{matsko99prl}
A.~B. Matsko, V.~V. Kozlov, and M.~O. Scully, ``{Backaction Cancellation in
  Quantum Nondemolition Measurement of Optical Solitons},'' {\em Phys. Rev.
  Lett.}, vol.~82, p.~3244–3247, Apr 1999.

\bibitem{80a1BrThVo}
V.~Braginsky, Y.~Vorontsov, and K.~Thorne, ``{Quantum Nondemolition
  Measurements},'' {\em Science}, vol.~209, p.~547–557, 1980.

\bibitem{81a1BrVoKh}
V.~Braginsky, Yu.I.Vorontsov, and F.~Y. Khalili {\em Sov. Phys. — JETP
  Lett.}, vol.~33, p.~405, 1981.

\bibitem{Clerk08}
A.~Clerk, F.~Marquardt, and K.~Jacobs, ``{Back-action evasion and squeezing of
  a mechanical resonator using a cavity detector},'' {\em New Journal of
  Physics}, vol.~10, p.~095010, 2008.

\bibitem{Clerk2015}
E.~Wollman, C.~Lei, A.~Weinstein, J.~Suh, A.~Kronwald, F.~Marquardt, A.~Clerk,
  and K.~Schwab, ``{Quantum squeezing of motion in a mechanical resonator},''
  {\em Science}, vol.~349, no.~6251, p.~952–955, 2015.

\bibitem{Pirkkalainen2015}
J.-M. Pirkkalainen, E.~Damskagg, M.~Brandt, F.~Massel, and M.~Sillanpaa {\em
  Phys. Rev. Lett.}, vol.~115, p.~243601, 2015.

\bibitem{Buchmann2016}
L.~Buchmann, S.~Schreppler, J.~Kohler, N.~Spethmann, and D.~Stamper-Kurn,
  ``{Complex Squeezing and Force Measurement Beyond the Standard Quantum
  Limit},'' {\em Physical Review Letters}, vol.~117, p.~030801, 2016.

\bibitem{18a1VyMaJOSA}
S.~Vyatchanin and A.~Matsko, ``{On sensitivity limitations of a dichromatic
  optical detection of a classical mechanical force},'' {\em Journal of Optical
  Siciety of America B}, vol.~35, p.~1970–1978, 2018.

\bibitem{Povinelli05ol}
M.~L. Povinelli, M.~Lončar, M.~Ibanescu, E.~J. Smythe, S.~G. Johnson,
  F.~Capasso, and J.~D. Joannopoulos, ``{Evanescent-wave bonding between
  optical waveguides},'' {\em Opt. Lett.}, vol.~30, p.~3042–3044, Nov 2005.

\bibitem{maslov13pra}
A.~V. Maslov, V.~N. Astratov, and M.~I. Bakunov, ``{Resonant propulsion of a
  microparticle by a surface wave},'' {\em Phys. Rev. A}, vol.~87, p.~053848,
  May 2013.

\bibitem{96a1VyMaJETP}
{S.P. Vyatchanin} and {A.B. Matsko}, ``{Quantum variation scheme of measurement
  of force and compensation of back action},'' {\em Sov. Phys. – JETP},
  vol.~82, p.~107, 1996.

\bibitem{matsko97apb}
{A.B. Matsko} and {S.P. Vyatchanin}, ``{A ponderomotive scheme for QND
  measurement of quadrature component},'' {\em Applied Physics B}, vol.~64,
  no.~2, p.~167–171, 1997.

\bibitem{LiPRA2019}
X.~Li, M.~Korobko, Y.~Ma, R.~Schnabel, and Y.~Chen, ``{Coherent coupling
  completing an unambiguous optomechanical classification framework},'' {\em
  Physical Review A}, vol.~100, p.~053855, 2019.

\bibitem{Walls2008}
D.~Walls and G.~Milburn, {\em {Quantum optics}}.
\newblock Springer-V Berlin Heidelbergerlag, 2008.

\end{thebibliography}

\end{document}